\documentclass[journal=jacsat,manuscript=article]{achemso}
\usepackage[version=3]{mhchem} 
\usepackage{chemfig}
\usepackage{xcolor}
\usepackage{amssymb}
\usepackage{tabularx} 
\author{Amaury Coste}
\affiliation{Univ. Grenoble-Alpes, CEA, IRIG-MEM-L Sim, 38000 Grenoble, France}
\email{amaury.coste@cea.fr}
\author{Thomas Meyer}
\affiliation{Univ. Grenoble Alpes, Univ. Savoie Mont Blanc, CNRS, Grenoble INP, LEPMI, 38000 Grenoble, France}
\author{Claire Villevielle}
\affiliation{Univ. Grenoble Alpes, Univ. Savoie Mont Blanc, CNRS, Grenoble INP, LEPMI, 38000 Grenoble, France}
\author{Fannie Alloin}
\affiliation{Univ. Grenoble Alpes, Univ. Savoie Mont Blanc, CNRS, Grenoble INP, LEPMI, 38000 Grenoble, France}
\author{Stefano Mossa}
\affiliation{Univ. Grenoble-Alpes, CEA, IRIG-MEM-L Sim, 38000 Grenoble, France}
\author{Benoit Coasne}
\affiliation{ Universit\'e Grenoble Alpes, CNRS, LIPhy, Grenoble, 38000, France}
\alsoaffiliation{ Institut Laue Langevin, 38042 Grenoble,
France}
\email{benoit.coasne@univ-grenoble-alpes.fr}
\title{Interplay of Structure and Dynamics in Solid Polymer Electrolytes: a Molecular Dynamics Study of LiPF$_6$/polypropylene carbonate}
\begin{document}

%
\begin{abstract}

Solid-state batteries (SSB) are emerging as the next generation of electrochemical energy storage devices. In this context, obtaining  high energy density batteries relies on the use of solid polymer electrolytes (SPE) that are electrochemically stable with respect to lithium metal and  high potential positive electrode (both conditions being difficult to achieve without chemical degradation). Here, molecular dynamics simulations are used to investigate the interplay of structure and dynamics of carbonate-based SPE made up of polypropylene carbonate and lithium hexafluorophosphate (LiPF$_6$) at salt concentrations ranging from 0.32 to 1.21 mol/kg. On the one hand, the structural properties of such SPE are studied under ambient pressure and at the experimentally relevant temperature $T = 353$ K. On the other hand, considering that the very slow processes involved in these systems are out-of-reach of molecular dynamics, the dynamic properties are simulated at high temperature up to 900~K and then extrapolated to $T = 353$ K using Arrhenius' law. Our results reveal strong ionic correlations with a limited fraction of free ions and a prevalence of negatively charged clusters (particularly at the highest salt concentrations). The self-diffusion coefficient of Li$^+$ exceeds that of PF$_6^-$ at high temperature due to the weaker Li$^+$-carbonate and ion-ion interactions. However, the Li$^+$ mobility at $T = 353$ K is lower than that of the anion, in agreement with the typical experimental SPE behavior reported in the literature. As expected, our MD simulations show that the ionic conductivity $\sigma$ increases with temperature. Moreover, $\sigma$ at $T = 353$ K  exhibits a maximum at a salt concentration between 1.0 and 1.1~mol/kg. Overall, our estimated physico-chemical parameters indicate that strong ion correlations can be optimized to design better SPE. In this context, the Arrhenius extrapolation approach employed here provides insights into ion transport mechanisms in SPE.
\end{abstract}

\section{1. Introduction}

With the increasing demand for high-energy-density devices, solid-state batteries (SSB) are considered promising for electrochemical energy storage~\cite{dunn2011,bresser2018,Titirici_2024}. Compared to conventional lithium-ion batteries that rely on a liquid electrolyte, SSB provide (1) higher energy density through the use of lithium metal as negative electrode and (2) enhanced safety by suppressing  flammable organic solvent~\cite{zhou2016,dong2024,ngo2025}. In practice, solid electrolytes are classified in two main categories: solid polymer electrolytes (SPE) and ceramic electrolytes (CE). Due to their brittle nature, CE are difficult to manufacture -- especially considering that they need to be densified to be used (therefore causing issue for electrode preparation where intimate contact must be achieved between the active material, the conductive additive, and the solid electrolyte). In contrast, SPE are cheap and flexible materials which are easy to manufacture and can be mixed easily with other electrode components~\cite{Song2023nat,li2023,2023polreview}.  As a result, features of the polymer material in SPE [e.g. poly(ethylene oxide) polymer (PEO)] have been investigated since the 1970's and PEO-based SPE have been commercialized~\cite{xue2015,thiam2019,barbosa2022,song2023,LI2023_intro}.

However, SPE exhibits low ionic conductivities at room temperature due to PEO crystallinity and slow dynamics with this high molecular weight polymer~\cite{daems2024}. In addition, due to the weak stability of the ether function against oxidation (which occurs at 3.7 V \textit{vs.} Li$^+$/Li), it is not possible to use high voltage electrode materials such as those of the NMC family [Li$_x$Ni$_a$Mn$_b$Co$_c$O$_2$, with (a+b+c=1)]. In fact, only a limited number of coordinating polymers display good electrochemical stability at high voltage. Chemically, these polymer electrolytes should contain polar groups such as carbonate or nitrile functions~\cite{Cheng2021}. In this context, polypropylene carbonate polymers (PPC) have proven to be a valid alternative to PEO~\cite{zhang2015,zhao2016,yue2018,xiao2023,Sashmitha2023,foran2024,zhang2024}. In fact, it is a low-cost and biodegradable polymer~\cite{du2004} with a chemical structure which closely resembles that of conventional carbonate-based electrolytes (therefore implying good compatibility with lithium salts and favorable interfacial contact with conventional electrodes)~\cite{yu2010,zhou2013}. Unfortunately, these polymers have high glass transition temperatures $T_g$ leading to slow ionic/polymer mobility and conductivity at room temperature~\cite{2023polreview}.

Optimizing the electrochemical performance of SPE-based batteries requires enhancing the transport of ionic species which, in turn, necessitates a detailed understanding of the system’s structural and dynamic properties~\cite{Yang2022,Yu2025}. While extensive studies using all-atom~\cite{Gudla2021,mabuchi2021} and coarse-grained~\cite{shen2020,shen2020b,Rajahmundry2024} Molecular Dynamics simulations have focused on the structural and dynamic properties of PEO/LiTFSI-based SPEs, PC-based SPEs have received significantly less attention~\cite{Gerlitz2023}. To address this knowledge gap, we use here molecular modeling to investigate the interplay between the structure and dynamics of  LiPF$_6$/PPC-based SPE as a function of the salt concentration $c_\textrm{s}$. 
In more detail, to provide a comprehensive picture of the underlying molecular mechanisms at play, we simulate SPE at different salt concentrations $c_{\rm s}$ from 0.32 to 1.21~mol/kg (which correspond to carbonate/salt ratios  $R_\textrm{C/Li}$ varying from 30 to 8). To shed light on the interplay between structural and dynamical properties in LiPF$_6$/PPC-based SPE, we investigate the properties of the polymer matrix, ionic local coordination shell, and the dissociation state of the salt in the electrolyte (free ions, ion pairs, ion clusters) as a function of the salt concentration $c_{\rm s}$. The dynamic and transport properties -- including the ion diffusion coefficient $D_{\rm s}$ and conductivity $\sigma$ -- are determined at several high temperatures $T$. Arrhenius' law is then employed to extrapolate the behavior at low temperatures relevant to the practical use of SPE. Based on these data, we provide an atom-scale description of the structure and dynamics of PPC-based SPE as a function of the salt concentration/temperature, which can contribute to the design of high-performance solid electrolytes.

\section{2. Experimental and Computational Methods}

\section{2.1. Experimental Section}

To validate the relevance of the atomistic model employed in MD, we measured the density of several PPC-based SPEs at different salt concentrations $c_{\rm s}$. In what follows, we provide details on the polymer electrolyte elaboration and the density measurements performed.

\noindent \textbf{Polymer electrolyte elaboration.}
The polymer electrolytes were prepared in an argon-filled glovebox (H$_2$O and O$_2$ concentration $< 0.1$ ppm).  Polypropylene carbonate -- hereafter called PPC, (Mn = 50,000 by GPC, Sigma-Aldrich, stored at -20 $^{\circ}$C) -- was used as received while lithium hexafluorophosphate LiPF$_6$ (Tokyo Chemical Industry, purity $>$ 97.0$\%$, stored under Ar) was used after drying for 2 days under dynamic vacuum at 110 $^{\circ}$C. Three polymer electrolytes with various lithium contents were prepared together with a reference material composed of pure PPC. The polymer and the lithium salt were dissolved in acetonitrile (Sigma-Aldrich, anhydrous, purity 99.8$\%$). After 2 hours of magnetic stirring at 200 rpm at a temperature of 80$^{\circ}$C, homogeneous solutions were cast in Petri dishes. The samples were then dried overnight at 60$^{\circ}$C under dynamic vacuum. 

\noindent \textbf{Density Measurements.} To estimate the density of each sample, a Mettler Toledo 33360 density determination kit was used. The kit is based on the Archimedes principle, which is particularly useful for irregularly shaped solids such as polymers. Here, we employed heptane (Sigma-Aldrich, purity~$\geqslant$~99.99\%) that does not dissolve or swell the polymer electrolyte. The experiments were carried out at room temperature in an argon-filled glovebox (${\rm H_2O}$ and ${\rm O_2}<$ 0.1 ppm).

\section{2.2. Molecular Simulation Section}

\begin{figure}[t]
\centering
\includegraphics[width=0.5\linewidth]{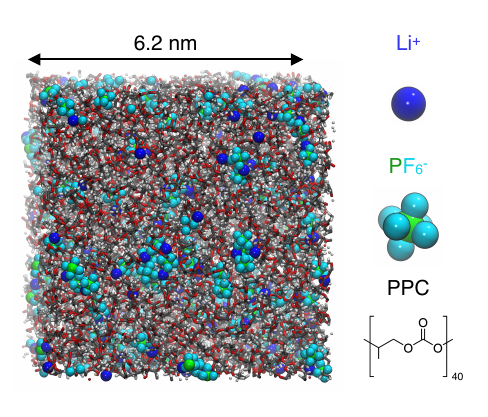}
\caption{Typical molecular configuration of the molecular model of LiPF$_6$/PPC SPE. The Li$^+$ cations are shown as blue spheres while the PF$_6^-$ anions are shown as green and cyan spheres. The PPC chains are made up of 40 monomers where the carbon, oxygen and hydrogen atoms are shown as gray, red and white spheres, respectively. The monomer chemical composition and structure are shown in the bottom right of the figure.}
\label{fig:1}
\end{figure}

\noindent \textbf{SPE molecular model.}  Fig.~\ref{fig:1} shows a typical molecular configuration of our LiPF$_6$/PPC model together with a zoomed view of the different chemical species that form this SPE. Each simulation box contains PPC chains made up of 40 monomers, therefore leading to a molecular weight of 4,100 g/mol. Although this value is about 120 times smaller than that of the average experimental polymer chain, we believe our polymer chains are sufficiently long to behave like the experimental system on time scales considered in MD. In practice, the polymer lengths considered here are similar to those employed in other MD simulations on PPC-based SPE~\cite{Gerlitz2023}. The ends of the chains were symmetrised and terminated with isopropyl groups  CH(CH$_3$)$_2$. The electrolyte compositions considered in the present work are reported in Table~\ref{tbl:c_choosed}. In practice, we modeled five SPEs together with the pure polymer. Note that the number of PPC chains (and, therefore, the total number of carbonate groups) remains constant in all cases, while the number of ion pairs varies to match the target salt concentration $c_\textrm{s}$. 
In addition, to analyze the role of the electrostatic interactions and of the polymer matrix on ion mobility, we simulated both the pure polymer melt and the electrolyte mixture with $R_{\rm C/Li}=$~8 for a system formed by substantially shorter chains (comprising only three monomers). In any case, the total number of carbonate groups is about the same as that for the long chain case so that the comparison between short and long chains is made for the same ionic concentration $c_\textrm{s}$.

Bonded and non-bonded interactions for the polymer were described using the OPLS-AA force field~\cite{opls} with partial atomic charges taken from Ref.~\cite{silva}. The Li$^{+}$ and PF$_6^-$ ions were described using the CL\&P force field~\cite{Padua2004,padua2004b}. With this force field, geometric mixing rules are applied to determine the Lennard-Jones parameters for the polymer/polymer interactions, whereas arithmetic mixing rules are employed for all other cross-interactions. The partial charges of the ions were scaled by a factor of 0.8 to account in an effective fashion for polarization effects and improve the description of ion dynamics as demonstrated in previous works~\cite{Costa2015,Mogurampelly2017,Gerlitz2023}. All molecular simulations were performed with periodic boundary conditions applied in the $x$, $y$, and $z$ directions. Non-bonded interactions were calculated within a cutoff distance $r_{\rm c}=12 $~\AA. Long-range electrostatic interactions were evaluated with the PPPM solver~\cite{pppm}. 
All simulations were carried out with LAMMPS~\cite{LAMMPS,BROWN2011898,BROWN2012449}. Newton's equation of motion was integrated using the velocity-Verlet algorithm~\cite{Verlet1, Verlet2} with a time step $\delta t=1$~fs. At $T = 900$ K, we used a shorter time step $\delta t=$~0.5 fs. Note that in all cases the C-H bonds were constrained using the SHAKE algorithm~\cite{shake}. We employed the Nos\'e-Hoover thermostat~\cite{nosehoover} with a coupling constant $\tau_T = 0.1$~ps. For the NPT simulations at $P=1.0$~bar, we additionally coupled the systems to a Nos\'e-Hoover barostat with a relaxation time $\tau_P =$~1.0~ps.

\begin{table}[t]
\caption{Composition of the LiPF$_6$/PPC  SPE considered in Molecular Dynamics. 
For every salt concentration $c_{\rm s}$, the carbonate/salt $R_\textrm{ C/Li}$, the number of LiPF$_6$ ions $N_{\rm salt}$ and the number polymer chains $N_{\rm pol.}$ for the long (40 monomers) and short (3 monomers) chains are reported.
}
\label{tbl:c_choosed}
\begin{tabular}{ccccc}
\hline\hline
 &$c_{\rm s}$ (mol/kg)&$R_{ C/Li}$  &$N_{\rm salt}$&$N_{\rm pol.}$\\
\hline\hline
long chain &-& - & - &40\\
&0.32 & 30 &53&40\\
& 0.48 &20 &80&40\\
&0.61& 16&100&40\\
&0.81&12&133&40\\
& 1.21 &8&200&40\\
&&&&\\
short chain & - & - &-&533\\
& 1.21 & 8 &200&533\\
\hline\hline
\end{tabular}
\end{table}

\noindent \textbf{Preparation protocol.} In order to prepare initial configurations for the SPE listed in Table~\ref{tbl:c_choosed}, we employed the same thermodynamic path in a systematic fashion. In brief, we prepared the initial configuration using a 200 ps NVT run at constant volume and a temperature of $T=$1500~K with a box length $L=$~60 nm. At this very low density, the system consists of an ideal gas of polymer chains and ions (in this initial phase, the charge of the salt ions were set to zero to avoid ion pairing/precipitation). Then, starting from this initial configuration, we performed an NPT rum at $T =$ 1500 K and $P=$~1~bar for 1~ns to condense the system and reach the volume corresponding to ambient pressure. This condensation step was followed by a 500~ps annealing from $T=$ 1500 K to 1000 K at constant volume $V$. The system was then further condensed at $T = 1000$ K at constant atmospheric pressure for 1~ns to eventually reach a density of about 0.9-1.0 g/cm$^3$ (followed by an additional 1~ns annealing from $T = 1000$ K to 700 K at constant volume and a longer relaxation at constant volume over 5~ns at the latter temperature). Finally, we further reduce the temperature from 700 to 353 K during 5~ns at constant ambient pressure $P = 1$ bar. At this point, the ion partial charges were set to their actual values and the system was equilibrated at $T = 353$ K and $P=1$~bar for 40 to 50~ns to reach the final constant density.

We note that the relaxation time of the PPC-based SPE at $T = 353$ K is much longer than the timescales accessible with molecular dynamics simulations since the system's viscosity is typically of the order of a few MPa$\cdot$s~\cite{wagner2023}. As a consequence, the dynamics that we can reasonably probe/access at room temperature mostly consists of localised rearrangements of the polymer matrix. To reasonably quantify the transport properties at each salt concentration, we therefore performed molecular simulations at higher temperatures: $T=$~600, 650, 700, 800, and 900~K. In practice, we started from $T = 353$ K and applied sequential temperature ramps of 15 ns each to reach the target temperature. Each ramp was conducted at constant volume, therefore resulting in a temperature increase along isochores corresponding to the density of the ambient system at $T = 353$ K for each concentration. At every temperature, we eventually performed a long equilibration in the NVT ensemble for 50 ns followed by the production run sufficiently extended to reach the diffusive (i.e., Fickian) regime for all systems. 

\section{3. Results}

\subsection{3.1. Structural properties}

\begin{figure*}[t]
\centering
\includegraphics[width=0.95\linewidth]{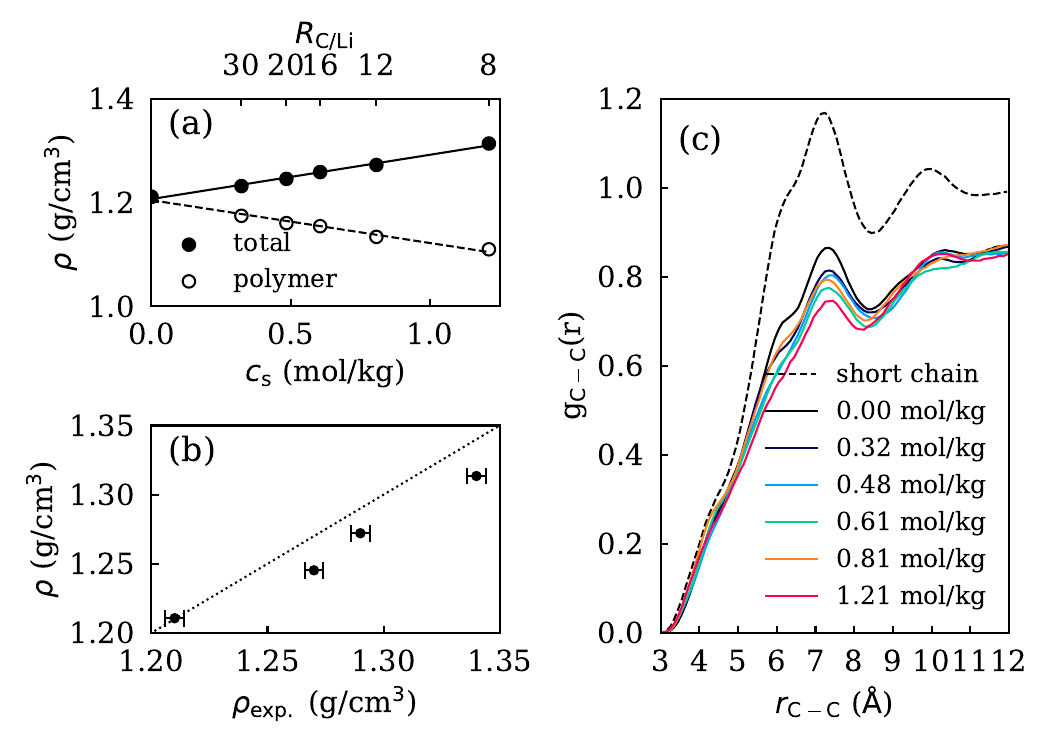}
\caption{(a) Computed total density $\rho$ and polymer matrix density $\rho_{\rm p}$ (without considering the salt mass) as a function of the salt concentration $c_\textrm{s}$. (b) Computed \textit{versus} experimental densities $\rho$ for the different SPEs considered in this study. The dashed line indicates the condition $\rho = \rho_\textrm{exp.}$ (c) Intermolecular radial distribution function $g_{\rm C-C}(r)$ between the C atoms of the carbonate function for the short chain polymer $n=3$ without salt (black dashed line), the  long chain polymer $n=40$ without salt (black solid line) and SPEs with different salt concentrations (solid color lines). 
\label{fig:2}}
\end{figure*}

\noindent \textbf{Polymer nanostructure.}
We determined the structural properties of the system at $T = 353$ K and at ambient pressure using MD trajectories of 60~ns. Fig.~\ref{fig:2}(a) shows both the total electrolyte mass density $\rho$ and that of the polymer matrix $\rho_{\rm p}$ as a function of the salt concentration $c_{\rm s}$. A linear increase in the SPE density is observed upon increasing $c_{\rm s}$, which is mirrored by an opposite trend for $\rho_{\rm p}$. This result indicates that the addition of salt increases the volume occupied by the polymer while increasing the overall density of the system because of the added salt ions. The accuracy of our modeling can be assessed by comparing the computed total density with the experimental results obtained as detailed in the experimental methods section. Fig.~\ref{fig:2}(b) shows both data sets for the pure polymer and SPE with three different concentrations [$c_{\rm s}=$~0.48, 0.81, and 1.21~mol/kg]. Although our simulations slightly underestimate the experimental density at all concentrations, the linear trend observed experimentally is reproduced quite satisfactorily using  molecular dynamics.  

We also assessed the details of the polymer nanostructure by computing the partial radial distribution functions defined as
\begin{equation}
g_{\alpha \beta}(r)=\frac{V}{N_{\alpha} N_{\beta}} \sum_{i=1}^{N_{\alpha}} \sum_{j=1}^{N_{\beta}}\left\langle\delta\left(r-\left|\vec{r}_{j}-\vec{r}_{i}\right|\right)\right\rangle,
\end{equation}
where $N_{\alpha}$ and $N_{\beta}$ are the numbers of atoms of type $\alpha$ and $\beta$ while $\vec{r}_{i}$ and $\vec{r}_{j}$ are the positions of atoms $i$ and $j$. Fig~\ref{fig:2}(c) shows $g_{\rm CC}(r)$ for the carbon atom of the carbonate group at different concentrations $c_\textrm{s}$ (intramolecular interactions between carbon atoms of the same chain were excluded). We also report the polymer structure for the short-chain polymer ($n=3$) without salt  and for the  long-chain polymer  ($n=40$) without salt. The radial distribution function $g_{\rm CC}(r)$ for short polymer chains without salt (black dashed line) exhibits a well-defined main peak at 7~\AA{} followed by a minimum at 8.5~\AA{} and a second peak at 10~\AA{} (such nanometric structure is similar to what is typically observed for liquid solvents). For the long-chain polymer without salt, the radial distribution function $g_{\rm CC}(r)$ (solid black line) displays peaks at the same positions, although their amplitude decreases due to polymer entanglement.

Despite small variations in the peak amplitudes, the radial distribution function $g_{\rm CC}(r)$ does not show any shift in the peak positions upon changing the concentration $c_{\rm s}$. This finding can be attributed to the small size of the salt ions and the low to moderate salt concentrations considered in the present study. Note that we prepared the systems at each concentration independently of each other to minimize correlations among the different numerical samples. 
The absence of marked impact of the concentration $c_\textrm{s}$ on the polymer nanostructure was confirmed by calculating in Table~\ref{tbl:pol_carac} the gyration radius $R_g$, the head-to-tail distance $l_{\rm h-t}$, and the interchain carbon-carbon contact number $N_{inter-chain}$ as a function of $c_{\rm s}$. On average, $R_g \sim 12$ ~\AA{} and $l_{\rm h-t} \sim 35$~\AA{} show no pronounced dependence on $c_{\rm s}$. On the other hand, $N_{inter-chain}$ decreases slightly but in a monotonous fashion with $c_{\rm s}$ due to the insertion of more and more ions between the polymer chains. Altogether, these results show that the addition of salt -- at least in the range of concentrations explored in this study -- does not induce significant modifications of the polymer structure.

\begin{table*}[t]
\caption{Average and standard deviation (in parentheses) of the polymer gyration radius $R_g$, head to tail distance $l_{\rm h-t}$, and interchain contact number $N_{inter-chain}$ for the SPE at $T = 353.15$ K and different concentrations $c_{\rm s}$.
}
\label{tbl:pol_carac}
\begin{tabular}{ccccc}
\hline\hline
$c_{\rm s}$ (mol/kg)&$R_{\rm C/Li}$ & $R_g$ (\AA)& $l_{\rm h-t}$ (\AA)&$N_{inter-chain}$ \\
\hline\hline
0.00&-&12.0 (3.0)&  34.3 (13.2)&3.1 (0.22)\\
0.32&30&11.4 (2.5)&32.4 (13.7)&2.9 (0.25)\\
0.48&20&11.8 (2.7)&33.4 (11.6)&2.8 (0.22) \\
0.61&16&11.9 (4.0)&34.6 (16.6)&2.7 (0.23)\\
0.81&12&12.2 (3.0)&37.5 (11.6)&2.7 (0.23)\\
1.21 &8&11.9 (3.0)&36.1 (10.8)&2.6 (0.25)\\
\hline\hline
\end{tabular}
\end{table*}

\begin{figure}[h!]
\centering
\includegraphics[width=0.5\linewidth]{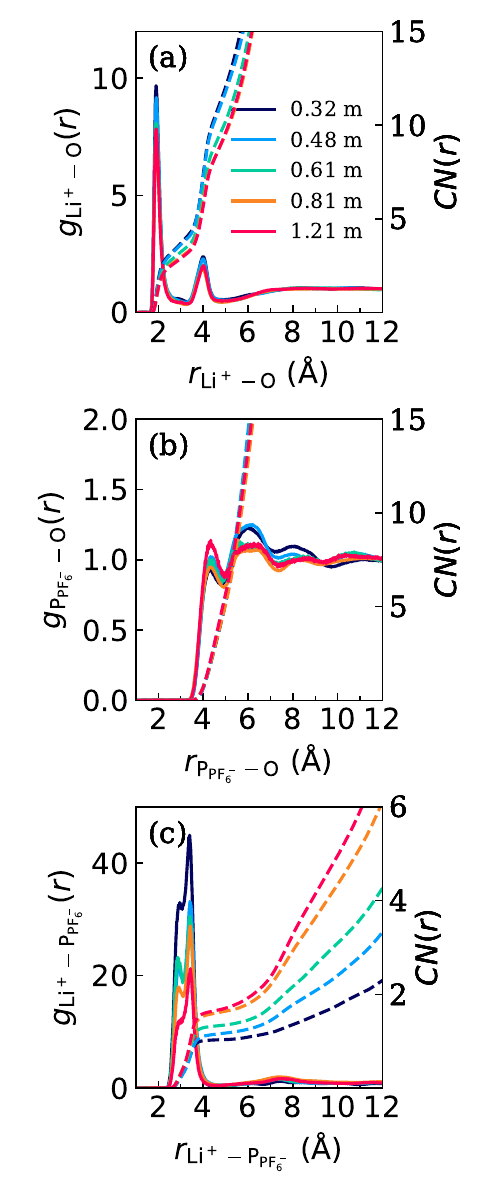}
\caption{Partial radial distribution function $g(r)$ (solid lines) and associated coordination number $CN(r)$ (dashed lines) at $T = 353$~K: (a) Li$^+$-O, (b) P$_{\rm PF_6^-}$-O and (c) Li$^+$-P$_{\rm PF_6^-}$.}
\label{fig:3}
\end{figure}

\noindent \textbf{Ion coordination.} The structural features of the ions (Li$^+$ and PF$_6^-$) were also analyzed using partial radial distribution functions calculated at $T = 353$~K. The $g_{\alpha\beta}(r)$ functions (left $y$-axis) together with the associated coordination numbers $CN_{\alpha\beta}(r)$ (right $y$-axis) are shown in Fig.~\ref{fig:3} for $\alpha={\rm Li}$/$\beta=\rm O$, $\alpha={\rm P_{PF_6^-}}$/$\beta=\rm O$, and $\alpha={\rm Li}$/$\beta={\rm P_{PF_6^-}}$ for different salt concentrations $c_\textrm{s}$.
For the cation-carbonate correlations in Fig.~\ref{fig:3}(a), we observe two peaks at distances of 1.93 ~\AA{} and 4.02~\AA{}, respectively. The first  peak, which corresponds to the formation of a closed packed coordination shell extending to $\simeq 3$~\AA{}, is in good agreement with previous {\it ab initio} molecular dynamics  results for LiTFSI/Propylene carbonate liquid electrolytes~\cite{Wang2024let}. Correspondingly, the coordination number at $3$~\AA{} decreases from 3.8 to 2.9 as the salt concentration $c_\textrm{s}$ increases from 0.32 to 1.21~mol/kg. This decrease is correlated with the increase in the number of ions in the solution, which leads to the formation of contact ion pairs and  ion clusters as will be discussed below. The second peak at $\simeq$~4.02~\AA{} arises from the interaction with oxygen atoms which belong to the interacting carbonate groups but are not directly in contact with the Li$^+$ cation. Again, the position of the peaks are not modified upon varying $c_{\rm s}$. The anion/carbonate correlations probed in Fig.~\ref{fig:3}(b) become non-negligible at distances larger than the radius of the cation/carbonate coordination shell with moderate peaks at 4.3~\AA{} and 6.0~\AA{}, respectively.  It has to be noticed that the amplitudes are close to 1. At all salt concentrations $c_\textrm{s}$, the corresponding coordination number is negligible for distances smaller than 4.3~\AA{} and then increases steeply up to $\simeq$~6 at the first minimum observed at 5.0~\AA{}. No modifications of the peak positions are observed with $c_{\rm s}$, while the amplitude slightly increases with the concentration $c_{\rm s}$.

Cation/anion structural correlations are shown in Fig.~\ref{fig:3}(c). Two partially overlapping peaks are observed at about 2.3 ~\AA{} and 3.44~\AA{} together with an additional weaker peak at around 7.4~\AA{}. These peaks correspond to two coordination modes: bidentate Li$^+$ (first peak) and monodentate Li$^+$ (second peak). The peak positions are in agreement with previous density functional theory (DFT) calculations~\cite{Borodin2001} for Li$^+$-PF$_6^-$ interactions in a poly(ethylene oxide) matrix.
The peak at 7.4~\AA{} corresponds to correlations of co-ions belonging to the same ionic cluster, which will be analysed in detail in the following. We also note that the coordination number $CN(r)$ in the first coordination shell increases from 1.0 to 1.7 as the salt concentration $c_\textrm{s}$ increases. A value larger than unity confirms that ions form both contact ion pairs and ionic clusters [Li$_x$(PF$_{6})_y]^{x-y}$. As expected, such ion pairs and clusters are more frequent at high salt concentrations, as will be discussed in detail below.

\begin{figure*}[tb]
\centering
\includegraphics[width=0.95\linewidth]{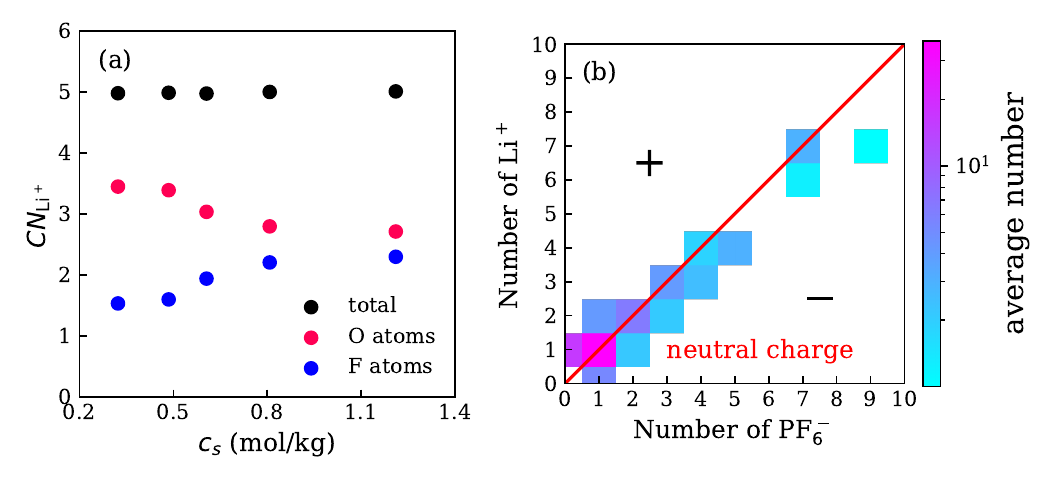}
\caption{(a) Coordination number of the first coordination shell of the Li$^+$ cations $CN_{\rm Li^+}$ as a function of $c_{\rm s}$. The total $CN_{\rm Li^+}$ and individual contributions with O and F atoms are shown in black, red and blue, respectively. The coordination shell is defined with a cutoff radius of 3.1~\AA{} based on the data shown in Figure~\ref{fig:2}.(a) and Fig. S1 of the Supporting Information. (b) Distribution of ion cluster [Li$_x$(PF$_{6})_y]^{x-y}$ for the SPE at $c_{\rm s} = 1.21$ mol/kg. Li$^+$ and PF$_6^-$ ions are considered to belong to the same cluster if they formed a contact ion pair CIP ($r_{\rm Li-P}~\leq$ 5.1~\AA{}). The red line represents equimolar (i.e., neutral) clusters where the number of cations and anions are equal.}
\label{fig:4}
\end{figure*}

A comprehensive view of the local coordination of Li$^+$ can be obtained by counting the number of oxygen (carbonate units) and fluorine (PF$_6^-$ anion) atoms comprised in spheres centered on Li$^+$ and with a radius $\simeq$~3.1~\AA  (corresponding to the position of the first minimum in the associated radial distribution functions, see Fig.~~\ref{fig:3}(a) and S1). Fig.~\ref{fig:4}(a) shows the salt concentration dependence of the total (black) coordination number together with the associated number of oxygen (red) and fluorine (blue) atoms contained in the sphere. The total $CN$ remains almost constant at about $\simeq$~5 upon increasing $c_{\rm s}$. However, the situation is different for the partial contributions as $CN$ for fluorine atoms increases while $CN$ for oxygen atoms decreases. This suggests that the anions gradually replace the oxygen atoms in the coordination shell of the cation upon increasing the salt concentration $c_\textrm{s}$. This finding has already been observed in other systems such as the Li$^+$ cation with different carbonate solvents -- typically at salt concentrations close to 1 mol/L. For ethylene carbonate, four oxygen atoms have been found in the first solvation shell using both classical MD~\cite{ravikumar2020,skarmoutsos2025} and DFT calculations~\cite{Skarmoutsos2015,Ponnuchamy2018}. Similarly, in propylene carbonate-based electrolytes, a value of 4.5 oxygen atoms have been reported in the first solvation shell of Li$^+$ cation using time-of-flight neutron experiments~\cite{Kameda2007}. In contrast, for dimethyl carbonate DMC, the first coordination shell of the Li$^+$ consists on average of 2.8 solvent molecules and 1.1 PF$_6^-$ anions~\cite{kameda2016}.

Our molecular simulation data also provide quantitative information about the extended ionic domains that form at high salt concentrations $c_{\rm s}$. To describe and quantify the association/dissociation states of the salt in SPE, we have used a graph approach to evaluate the global ionic connectivity network~\cite{vatin2021}. In practice, we associate (1) network nodes to the centers of mass of all ions and (2) network edges based on the distance criterium Li$^+$-P$_{\rm PF_6^-}~\leq5.1$~\AA (the latter corresponds to the boundary of the coordination shell as described above). This procedure allows us to determine the speciation of the ions in the SPE and to identify three different states for both ions: {\em i}) free (bare) ions (i.e. without any counterions in the first coordination shell); {\em ii}) contact ion pairs Li$^+$PF$_6^-$ (CIP); and {\em iii}) ion clusters made up of three or more ions. 
Fig.~\ref{fig:4}(b) shows a color map of the number of ion clusters with a given cation/anion composition occurring in each SPE (these data are averaged over the entire MD trajectory). Here, we consider the SPE with the largest salt concentration $c_{\rm s} =$ 1.21~mol/kg. The red solid line corresponds to the condition of neutral clusters (equal number of cations/anions) so that it delimitates  positively charged clusters (above the line) and negatively charged clusters (below the line). For the sake of clarity, considering that the simulation box contains 200 cations and 200 anions at this concentration, we only considered in this plot clusters and CIP with an occurrence of at least 1 per molecular configuration. 
These data show that CIP is the most common ionic form followed by free Li$^+$ and then small clusters of size ($\le$4). While the later can be positively charged, negatively charged, or neutral, large clusters tend to be neutral or negatively charged. It has to be noticed that the number of clusters is statistically small. However, the trend is consistent with previous observation obtained by means of classical molecular dynamics for LiTFSI/PEO-based SPE~\cite{molinari2018}. 
Table~\ref{table:cip} shows the  percentage of the three co-ion states for different $c_{\rm s}$. The same data obtained at higher temperatures are shown in Table~S1 of the Supporting Information. Only small fractions of free ions are observed for the SPE at the highest concentration $c_{\rm s}$. The fraction of free cations is about three times larger than that of free anions, therefore resulting in the formation of negatively charged clusters. The number of CIP is also significant as it represents almost 20~\% of the total salt content. As expected, upon lowering $c_{\rm s}$,  the fractions of CIP and free ions increase whereas the fraction of ion  clusters decreases. Also, we note that the overall speciation distribution does not show any substantial modification at higher temperatures as can be seen in Table~S1 of the Supporting Information.

\begin{table*}[tb]
\centering
\begin{tabularx}{\textwidth}{p{3.cm}*6{X}}   
\hline\hline
$c_{\rm s}$ (mol/kg) & \multicolumn{2}{c}{ Free single ions (\%)} & \multicolumn{2}{c}{ions in CIP (\%)} & \multicolumn{2}{c}{ions in cluster (\%)}\\
& \multicolumn{1}{c}{Li$^+$} & PF$_6^-$ &  \multicolumn{1}{c}{Li$^+$} & PF$_6^-$ &  \multicolumn{1}{c}{Li$^+$}  & PF$_6^-$\\
\hline\hline
0.32 & 24 & 15 & 43 & 43 & 33 & 42\\
0.48 & 15 & 6 & 43 & 43 & 42 & 51\\
0.61 & 16 & 5 & 31 & 31 & 53 & 64\\
0.81 & 13 & 3 & 23 & 23 & 64 & 74\\
1.21 & 8 & 3 & 18 & 18 & 74 & 79\\
\hline\hline
\end{tabularx}
\caption{Average distribution of Li$^+$ cations and PF$_6^-$ anions in three different states: free ions, LiPF$_6$ contact ion pairs (CIP), and ion clusters [Li$_x$(PF$_{6})_y]^{x-y}$. Free single ions are defined as ions with no counterion within the first cation/anion coordination shell. The other ions form either CIP or ion clusters.}
\label{table:cip}
\end{table*}

The large ionic association in SPE can be explained by estimating the dielectric constant $\epsilon$ of the polymer matrix from the dipole moment fluctuations~\cite{Neumann1983},
\begin{equation}
\epsilon/\epsilon_0=\frac{1}{3 V k_{\rm B}T}(\left< M^2\right> -\left< M\right>^2).
\end{equation}
We obtain $\epsilon = 1.3\;\epsilon_0$, which is quite small and of the order of magnitude of the experimental value $\epsilon= 3.0\; \epsilon_0$ at room temperature~\cite{Rieger2012,Rullyani2018}. This result contrasts with the case of cyclic carbonates, which exhibit a high dielectric constant -- $\epsilon \simeq$~65 and $\epsilon \simeq$ 95 for propylene and ethylene carbonate, respectively~\cite{payne1972,hall2015,zhang2021}. For the linear carbonate solvents and polymer carbonates (such as that considered here), the organization and solvation properties are different and the electrostatic screening of the charge is weaker -- therefore limiting salt dissociation. Also, the dielectric constant was studied by classical MD for liquid electrolytes at different temperatures and electrolyte compositions (of particular relevance to the present work, the system DMC/LiPF$_6$ was studied~\cite{Yao2021}). The results revealed similar $CN$ to those observed for our SPE. Furthermore, the DMC solvent exhibits a low dielectric constant ($\epsilon_{\rm MD }\simeq 1.44\epsilon_0$ and $\epsilon_{\rm exp }\simeq 3.17\epsilon_0$).

\subsection{3.2. Dynamics and transport}

\begin{figure*}[h!]
\centering
\includegraphics[width=0.95\linewidth]{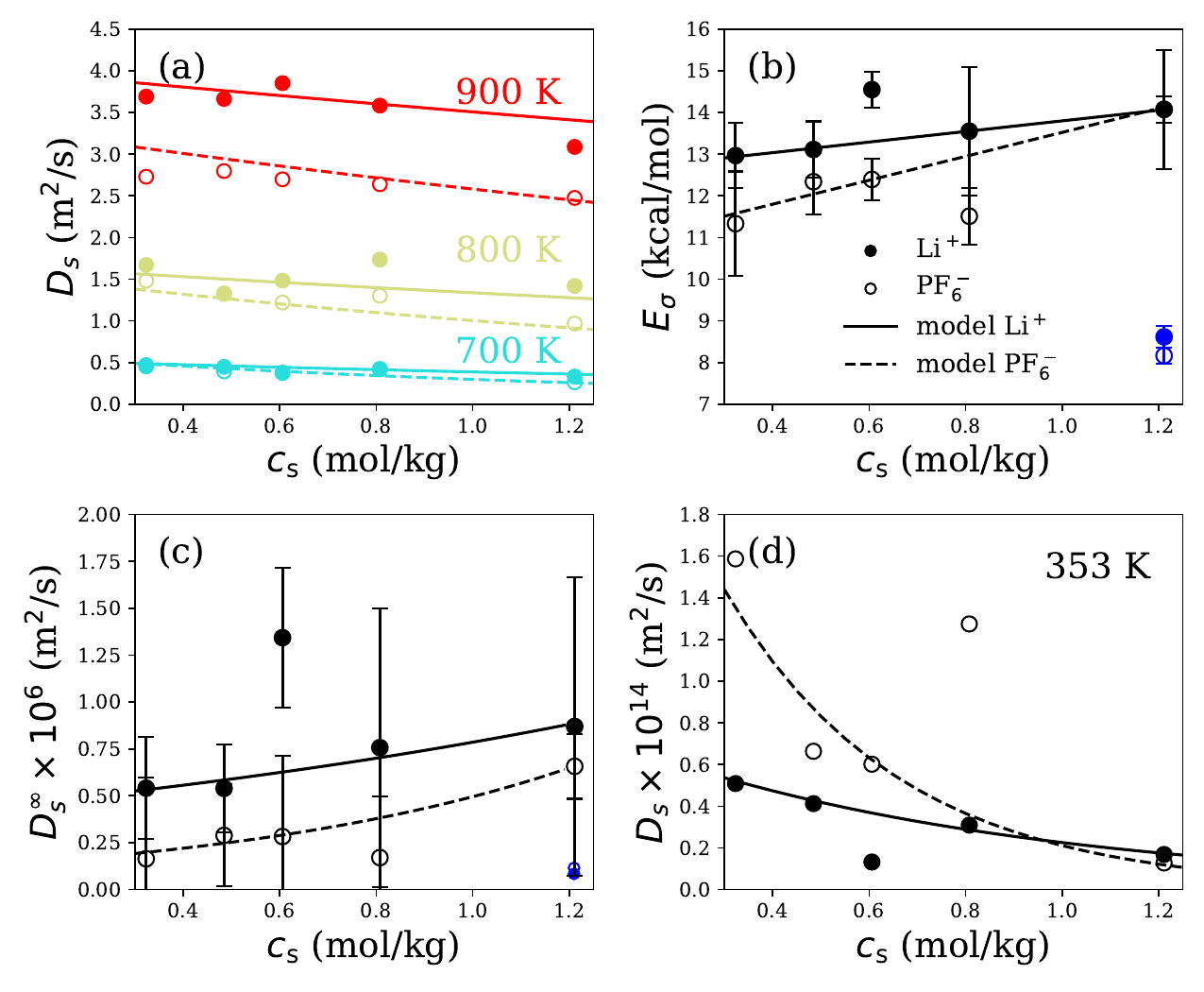}
\caption{(a) Self-diffusion coefficients $D_{\rm s}$ of Li$^+$ cation (closed circle) and PF$_6^-$ anion (open circle) as a function of $c_{\rm s}$. The data are shown at $T = 700$ K (cyan), $T = 800$ K (orange) and $T =  900$ K (red). (b) Activation energy $E_a$ for diffusion and (c) $D^{\infty}_{\rm s}$ calculated at every salt concentration $c_{\rm s}$ for the cations Li$^+$ (close circle) and the anions PF$_6^-$ (open circle). The dark blue data represent the values obtained when using SPE short chains. The line corresponds to the models employed for $E_a$ and $D^{\infty}_{\rm s}$ which are the parameters in Arrhenius' law. (d) The lines show the extrapolated self-diffusion coefficients at $T = 353$ K  as inferred from Arrhenius' law for Li$^+$ cations (closed circle) and PF$_6^-$ anions (open circle).}
\label{fig:5}
\end{figure*}

\noindent \textbf{Ionic Diffusion.} We calculated the self-diffusion coefficients $D_{\rm s}$ for the cations and anions at temperatures ranging from $T = 700$ K to 900 K. In more detail, using the Einstein relation, we extracted $D_{\rm s}$ from the ionic mean squared displacements $\left< \Delta r^2(t)\right>$ (Fig.~S3) in the long time limit (e.g. Fickian regime)~\cite{pranami2015,Kellouai2025}:
\begin{equation}
D_{\rm s} = \frac{1}{6} \lim_{t\to\infty} \frac{{\rm d}\left< \Delta r^2(t)\right>}{{\rm d} t},
\end{equation}
Fig.~\ref{fig:5}(a) shows $D_{\rm s}$ for the Li$^+$ cations (circle) and the PF$_6^-$ anions (open circle) as a function of $c_{\rm s}$ at $T=$ 700 K, 800 K and 900~K. As expected, increasing the temperature increases the diffusion of ions at all concentrations $c_{\rm s}$. For Li$^+$,  $D_{\rm s}$ is almost concentration independent in the range from $T =  700$ K to $T = 800$ K while it decreases as $c_{\rm s}$ increases at $T = 900$ K. In contrast, for PF$_6^-$, $D_{\rm s}$ decreases upon increasing $c_{\rm s}$ at all temperatures $T$. 

We can extrapolate these data using an Arrhenius law to predict the cation and anion diffusion coefficients at the experimental temperature $T=353$~K. As already explained, this approach allows us overcoming the fact that the relaxation time scale near room temperature exceeds those that can be reached using molecular dynamics. In practice, for each concentration $c_{\rm s}$, we have fitted our data at high temperatures against Arrhenius' law (Fig.~S4):
\begin{equation}
D_{\rm s}(c_{\rm s}, T) = D^{\infty}_{\rm s}(c_{\rm s}) \exp\left(-\frac{E_a(c_{\rm s})}{RT}\right),
\label{eq:arrhenius}
\end{equation}
where $D^{\infty}_{\rm s}$ is the diffusivity at infinite temperature -- that is where the thermal energy largely exceeds the activation energy for diffusion $E_a$.
Here, we note that the Vogel-Tammann-Fulcher (VTF) equation is often used to fit the transport properties of SPE because it takes into account the correlation between the ion transport and the segmental motion of the polymer~\cite{Diederichsen2017,aziz2018}. However, considering that our molecular simulations were conducted at temperatures significantly above the glass transition temperature, we found that Arrhenius' law accurately describes the physics at play for both ions as shown in Figs.~S4 and S5 of the Supporting Information. 

Fig.~\ref{fig:5}(b) shows  the concentration dependence of $E_{\rm a}$ for the cation Li$^+$ (red) and the anion PF$_6^-$ (black). For the Li$^+$ cation, $E_{\rm a}$ increases rather linearly with $c_{\rm s}$ as illustrated by the black line. However, we note that the overall increase in $E_{\rm a}$ is quite small as $E_{\rm a}$ remains very close to the average value of $\simeq$ 13.5 kcal/mol. In fact, taking $E_{\rm a}$ as a constant would not modify the conclusions drawn below. For the PF$_6^-$ anion, $E_{\rm a}$  increases linearly with a more pronounced variation with $c_{\rm s}$ from 11.3 kcal/mol at low salt concentration to 14.1 kcal/mol at high salt concentration. At the concentration $c_{\rm s} = $  1.21~mol/kg, $E_a$ is almost equal for both ions due to the low dissociation state for this concentration. In addition, we also show in the same figure  our results for the SPE at 1.21~mol/kg when the long polymer chains are replaced by short chains. In that case, $E_{\rm a}$ is significantly lower with a value of about $\simeq$ 9 kcal/mol for both ions. The variation of $E_{\rm a}$ is mainly due to the polymer mobility since the dissociation state and the dielectric constant for both systems are similar. Indeed, the fast dynamics of the short polymer chains necessarily decreases the activation energy and, hence, promotes ion diffusion and transport.
Fig.~\ref{fig:5}(c) shows the prefactor $D^{\infty}_{\rm s}$ for the cation Li$^+$ (in red) and the anion PF$_6^-$ (in black). In both cases, $D^{\infty}_{\rm s}$ increases with the salt concentration $c_\textrm{s}$. Like for $E_{\rm a}$, as a first-order approximation, we can fit $\ln D_s^{\infty}(c_{\rm s})$ against an exponential function with a linear argument. 

With our fits for $E_{\rm a}(c_{\rm s})$ and $D^{\infty}_{\rm s}(c_{\rm s})$, we can now predict the diffusivities of the ions at any concentration $c_{\rm s}$ and temperature $T$ using Eq.~(\ref{eq:arrhenius}). The results of this interpolation/extrapolation (solid and dashed lines) are found to be in reasonable agreement with our molecular simulation data (symbols) in Fig.~\ref{fig:5}(a). Using this procedure, we can extrapolate the values of $D_{\rm s}(c_{\rm s}, T)$ to experimentally relevant temperature conditions as shown in Fig.~\ref{fig:5}(d).
A few  observations are in order here. First, while Li$^+$ cation diffusion is faster than that of the PF$_6^-$ anion at high temperatures, an inversion in the self-diffusivity of the anion/cation occurs at low temperature so that the experimentally observed behaviour is recovered (slower cation diffusion near room temperature)~\cite{Timachova2015}. This result is due to the fact that ion/polymer correlations are less dominant at high temperatures, so that diffusion is mostly driven by the size of the ion (reciprocally, diffusion at low temperatures is mostly driven by ion/ion and ion/polymer interactions and is, therefore, less sensitive to the ion sizes). 
Second, while $D_{\rm s}$  for Li$^+$ is nearly independent of $c_{\rm s}$, $D_{\rm s}$ for PF$_6^-$ decreases significantly upon increasing salt concentration (from 1.5$\times10^{-14}$~m$^2/s$ at 0.32~mol/kg to 0.2$\times10^{-14}$~m$^2/s$ above 1.3~mol/kg). We note that $D_s$ at low $c_{\rm s}$ and its decrease could be overestimated by our linear model used to extrapolate the data at high temperatures to near room temperature. However, in any case, this behavior is in agreement with the experimental observations for the reference PEO/LiTFSI-based SPE  (e.g. ~\cite{Timachova2015}), where the effect of molecular weight and salt concentration on the ionic transport for the PEO/LiTFSI system was investigated by means of NMR. Overall, the data above indicate that -- although our PPC/LiPF$_6$ system behaves qualitatively in a fashion similar to the reference PEO/LiTFSI material -- it presents a lower ionic mobility in agreement with previous studies for PPC-based SPE~\cite{Gerlitz2023}.

\begin{figure*}[h!]
\centering
\includegraphics[width=0.95\linewidth]{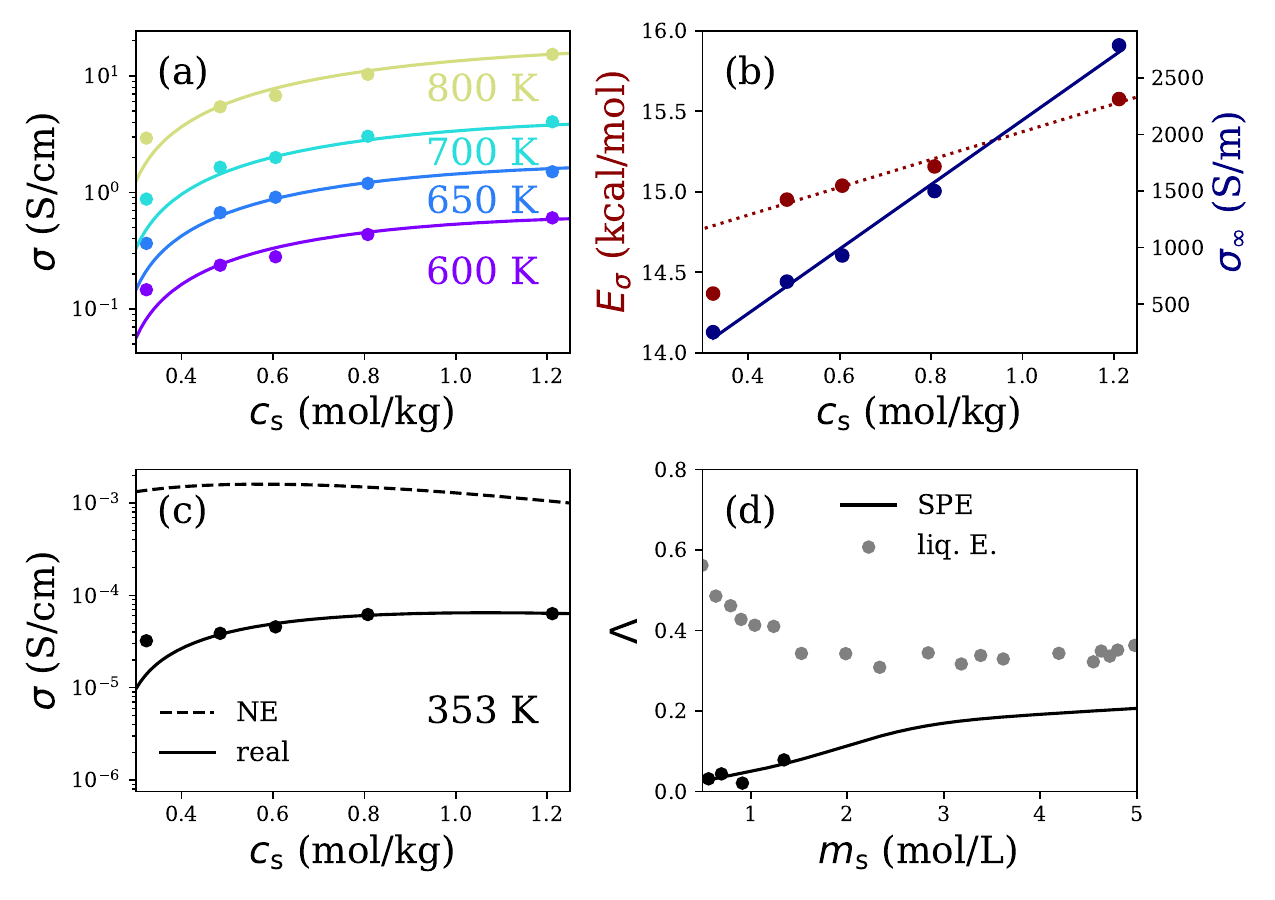}
\caption{(a) Ionic conductivity $\sigma$ as a function of the salt concentration $c_{\rm s}$ for $T  = 600$ K, 650 K, 700 K and 800~K. (b) Activation energy $E_{\sigma}$ (red) and $\sigma_{\infty}$ (blue) for the ionic conductivity $\sigma$ as a function of $c_{\rm s}$. The points corresponds to data while the lines are linear regressions.  (c) Real (straight line) and Nernst-Einstein (dashed line) ionic conductivities at $T = 353$ K as a function of $c_{\rm s}$ predicted from high temperature data using Arrhenius' law (see text). (d) ionicity $\Lambda$ at 353~K for the SPE (black) and for liquid electrolytes LiBF$_4$ in ethylene carbonate issued from ref~\cite{skarmoutsos2025}.}
\label{fig:6}
\end{figure*}

\noindent \textbf{Ionic conductivity.} In this section, we address how ion dynamics affects ionic conductivity $\sigma$ by varying the temperature $T$ and the salt concentration $c_{\rm s}$.
Following previous work~\cite{hartkamp2014,cazade2014,cox2019}, we performed out-of-equilibrium molecular dynamics simulations under an electric field at $T=$~600, 650, 700, and 800~K to determine the ionic conductivity $\sigma$. In short, an external electric field $\textbf{E}$ -- with $E =   |\textbf{E}|$ ranging from -0.03 to 0.03~V/\AA{} -- is applied along the $x$ direction and the resulting ionic current in the same direction $\textbf{J}_{{\rm ion}}(t)$ is monitored along the dipole following the dipole $\textbf{P}_{\rm ion}(t)$:
\begin{equation}
    \textbf{P}_{\rm ion}(t) =\frac{1}{V} \sum_{\rm ion} q_{\rm ion} x_{\rm ion}(t)
\end{equation}
As shown in Fig.~S5 and S6 of the Supporting Information,  $\textbf{J}_{{\rm ion}}$ data are generally quite accurate with very limited statistical noise. Fig.~\ref{fig:6}(a) shows the ionic conductivities $\sigma$ (points) obtained from the slope $\partial J_{{\rm ion},x}/\partial E$ as a function of concentration $c_{\rm s}$ for different temperatures $T$. 
As expected, for a given concentration $c_{\rm s}$, $\sigma$ increases upon increasing temperature $T$. Moreover, for a given temperature $T$, $\sigma$ increases with $c_{\rm s}$ as the density of charge carriers increases -- an effect that is enhanced upon increasing the temperature. In order to extrapolate these data to near-room temperature, as illustrated in Fig.~S7.(a) of the Supporting Information, we have fitted these ionic conductivity data with the following Arrhenius' law: 
\begin{equation}
\sigma_{\rm s}(c_{\rm s}, T) = \sigma^{\infty}_{\rm s}(c_{\rm s}) \exp\left(-\frac{E_{\sigma}(c_{\rm s})}{RT}\right).
\label{eq:arrheniusS}
\end{equation}
In practice, both the prefactor $\sigma^\infty$ and activation energy $E_{\rm \sigma}$ were determined for each concentration $c_{\rm s}$. We also note that the Arrhenius' fits are obtained with very limited statistical noise. Fig.~\ref{fig:6}(b) shows both the activation energies $E_{\rm \sigma}(c_{\rm s})$ (red circle) and $\sigma^\infty$ (blue circle) as a function of $c_{\rm s}$. $E_{\rm \sigma}$ increases by about 1~kcal/mol from 14.4 at $c_{\rm s} = 0.32$ mol/kg to 15.6~kcal/mol at $c_{\rm s} = 1.21$ mol/kg. 
This increase indicates stronger ion-ion interactions and cluster size as $c_{\rm s}$ increases. The simulated activation energies for conductivity are similar to the value of 15.6~kcal/mol obtained in Ref.~\cite{Chaurasia2015} for the semi-crystalline PEO+10 wt.$\%$ LiPF$_6$. For $\sigma^\infty$, we also observe a linear increase in $c_{\rm s}$ as the number of charge carriers. Both parameters were fitted with a linear regression function that represents well the trend observed in the range of $c_{\rm s}$ considered here. 

In order to estimate the impact of ionic correlations on ionic conductivity, we now compare the true ionic conductivity $\sigma$ to the Nernst-Einstein conductivity $\sigma_\textrm{NE}$, which can be assessed from the cation and anion diffusivities~\cite{Maginn_Messerly_Carlson_Roe_Elliot_2018,verma2024}:
\begin{equation}
\sigma_{\rm NE} =\frac{e^2}{V\,k_{\rm B}T}(N_+Z_{+}^2D_{S}^+ +N_{-}Z_{-}^2D_{S}^-).\label{eq:NE},
\end{equation}
where $V$ is the volume of the system, $N_+$ and $N_-$ are the numbers of cations and anions, and $Z_{+}e$ and $Z_{-}e$ their respective charge. $\sigma_{\rm NE}$ is considered to be the conductivity of charges that would diffuse without any ionic correlations so that $\sigma_\textrm{NE} \ge \sigma$ at all concentrations $c_\textrm{s}$. 
Fig.~\ref{fig:6}(c) shows $\sigma$ (solid line) and $\sigma_{\rm NE}$ (dashed line) as a function of $c_{\rm s}$ as extrapolated to a temperature of $T = 353$ K using Arrhenius' laws. The points represent the direct prediction for each SPE obtained with Arrhenius' law.
Interestingly, an ionic conductivity optimum of $\sim$~6.5$\times 10^{-5}$ S/cm is observed for a salt concentration between 1.0 and 1.1~mol/kg [see Fig.~S7(b) of the Supporting Information]. Although this conductivity maximum is in the salt concentration range expected experimentally (see for instance~\cite{skarmoutsos2025}), we note that here that it is a true prediction from our molecular simulation data obtained at high temperature. In particular,  for the range of salt concentrations under study, this optimum is not observed in the high-temperature data so that it really corresponds to a physical effect that only occurs at low to moderate temperatures [{\it e.g.} the decrease of the mobility of the ionic species due to ion/polymer and ion/ion interactions as illustrated in Fig.~S8(a) of the Supporting Information]. Furthermore, the conductivity $\sigma$ at $T=300$~K is one order of magnitude lower with a maximum of 1.3$\times 10^{-6}$ S/cm, which is similar to the experimental value reported for PPC-LiBOB SPE~\cite{Sashmitha2023}. For $\sigma_{\rm NE}$ at $T=353$~K, we also observe an optimum $\sim$~1.5$\times 10^{-3}$ S/cm around 0.6~mol/kg. Given these values, a factor of about 20 is observed between the Nernst-Einstein and real conductivities along with a shift in the optimum toward lower salt concentrations. Such differences can be explained by the fact that important ion-ion interactions in SPE (including ion pairing) are not taken into account in the Nernst-Einstein conductivity~\cite{Pang2021}. It was also demonstrated that the gap between both conductivities is enhanced by the viscosity of the system~\cite{shao2020v}.

Physico-chemical parameters are important to determine if an electrolyte is a good candidate for electrochemical applications. Among such quantities, the ionicity can be defined as the ratio of the true \textit{versus} the ideal conductivity.
\begin{equation}
\Lambda =\frac{\sigma}{\sigma_{\rm NE}}
\end{equation}
$\Lambda$  quantifies the degree of dynamic ion correlations in the system. Fig.~\ref{fig:6}(d) shows $\Lambda$ for our SPE at $T = 353$~K  along with a qualitative sketch illustrating the anticipated trend for liquid carbonate electrolytes. $\Lambda$ for the other temperatures of our SPE are plotted in Fig.~S8(b) of the Supporting Information. As expected, $\Lambda$ decreases as the temperature decreases at a given concentration $c_{\rm s}$ due to the stronger ion correlations in the SPE under near room temperature conditions. Interestingly, $\Lambda$ does not decrease with $c_{\rm s}$ as commonly observed for electrolytes due to the increase in the ion-ion correlations at large salt concentrations. For all temperatures, $\Lambda$ increases with $c_{\rm s}$ which departs from the common picture for electrolyte solutions~\cite{skarmoutsos2025}. We note that the increase between 0.5 and 1.21~mol/kg for all temperatures is very small ($<$ 0.1) (such  variations  are more pronounced for liquid electrolytes~\cite{skarmoutsos2025,Teherpuria2025}).  
$\Lambda$ value is less than 0.1 at near-room temperature, therefore suggesting that the Nernst-Einstein relation is not adequate to model the SPE transport properties. Such pronounced correlations are consistent with the structural analysis above, which showed that ions in SPE form a significant number of contact ion pairs and show a clear tendency to cluster even at the lowest concentration~\cite{lanord2019}. We also note that the ionicity $\Lambda$ is almost concentration independent in the range of 0.6-1.3~mol/kg for temperatures above $T = 700$ K, while it increases with $c_{\rm s}$ at higher temperatures. This is due to thermal motion at temperatures large enough such that $k_\textrm{B}T$ is comparable to or larger than the dissociation energy (therefore leading to more decorrelated ion transport). As a result, since the ionicity is defined as a per ion quantity, it becomes concentration independent provided that the ionic conductivity pertains to this decorrelated transport regime.

\section{4. Conclusion}

In this work, we investigated by means of Molecular Dynamics simulations the interplay between structure and dynamics in solid polymer electrolytes composed of polypropylene carbonate with different salt concentrations $c_{\rm s}$ of lithium hexafluorophosphate. Such PPC/LiPF$_6$-based SPE exhibit strong ion-ion interactions ({\it i.e.}, ion association) as revealed by the small fraction of free ions and the large fraction of ions involved in contact ion pairs and ion clusters. The number of free Li$^+$ cations slightly exceeds that of PF$_6^-$ anions, therefore leading to ionic clusters that tend to be neutral or negatively charged as their size increases (the negative charge of these clusters induces the mobility of Li$^+$ cation by migration in the opposite direction to the Li$^+$ concentration gradient induced by electrochemical reactions). This effect is further enhanced by the length of the polymer chain. Such SPE are low dielectric media so that electrostatic interactions are a key parameter to model these systems (as a result of only partial/inefficient coulomb screening). For the short chain polymer ($n=3$), the simulated $\epsilon_ \simeq~1.4\epsilon_0$ -- which is almost the value for the polymer made up of long chains ($n=40$). Consequently, the distribution of ions into the three different states -- single free ion, CIP and ion cluster  at 1.21~mol/kg -- follows the same trend for both the short-chain and long-chain polymers with variations attributed to the conformational constraints imposed by the extensive polymer chain length (we observed a somehow larger number of positive clusters of small size for the short chain SPE, Fig.~S2).

To assess ion transport at $T = 353$ K, where the relaxation time scale exceeds that accessible in typical molecular dynamics simulations, we employed a scaling based on Arrhenius' law.  We found that this method, which reproduces the expected behavior of SPEs at $T = 353$ K,  is a powerful tool that can be used for more complex setups (including electrolyte/electrode interfaces). Transport was first analysed by considering ionic self-diffusivities. While increasing the temperature leads to ion diffusion enhancement (especially that of Li$^+$ cation), the interaction Li$^+$/carbonate weakens with temperature so that it enhances the solvation/desolvation process of Li$^+$ cations. Also, we observed that $D_{\rm s}$ becomes concentration independent upon decreasing the temperature. 
The dynamics of the cations are strongly coupled to that of the segmental motion of the polymer -- as evidenced by the amplitudes of the main peak in the radial distribution functions $g_{\rm Li-O}(r)$. This marked interplay arises from the strong polymer/cation interactions. Anions, on the other hand, interact only weakly with the polymer matrix, therefore leading to faster diffusion compared to that of the cations at low $c_{\rm s }$. As $c_{\rm s }$ increases, the decrease in the self-diffusion coefficient of  PF$_6^-$ anion is attributed to ion pairing in these low-dielectric media as mentioned above. This interpretation is further supported by the amplitudes of the radial distribution functions $g_{\rm P-O}(r)$ and $g_{\rm P-Li}(r)$, which confirm that the anion interacts weakly with the polymer but quite strongly with the counterion.

Collective ion transport was also quantified by estimating the ionic conductivity $\sigma$. By extrapolating our high temperature data to $T = 353$ K, we found an optimum in the ionic conductivity at about $c_\textrm{s} = 1.1$ mol/kg. This conductivity optimum  shifts to higher $c_{\rm s}$ upon increasing $T$ as illustrated in Fig.~S8.(a) of the Supporting Information. The conductivity maximum, which is consistent with experimental data on liquid and polymer electrolytes (e.g. ~\cite{logan2018,Bolloli2015,skarmoutsos2025}), arises from the trade-off between (1) fast transport/small number of charge carriers at low salt concentrations  and (2) hindered transport/large  number of charge carriers at high salt concentrations. Upon increasing the salt concentration, the viscosity increases so that it reduces the ion mobility and then $\sigma$. Moreover, the dissociation state of the salt decreases upon increasing $c_{\rm s}$ due to the promotion of large cluster formations which can eventually lead to complete structural arrest~\cite{li2023}. Finally, the ionicity $\Lambda$ obtained throughout this study indicates limited transport performance of the investigated SPE. Indeed, $\Lambda <$~0.1 at $T = 353$ K suggests strong ion dynamical correlations and confirms that the Nernst-Einstein relation is not adequate to estimate transport properties. Overall, we find that LiPF$_6$/PPC SPE exhibit diminished transport properties compared to the reference SPE material made up of LiTFSI/PEO. Different strategies -- like the addition of plasticizing agents in the SPE (solvent, ionic liquid)~\cite{zhang2015,Gerlitz2023}, the elaboration of composite solid polymer electrolytes~\cite{zhu2020}, or the use of a binary salt mixture~\cite{li2021} -- could prove to be a valid and efficient strategy for the rational design of optimized electrolytes.

\begin{acknowledgement}
This work was supported by the Center of Excellence in Multifunctional Architectured Materials -- CEMAM (grant ANR-10-LABX-44-01) funded by the "Investments for the Future" Program, and by the French National Research Agency under the France 2030 program (Grant ANR-22-PEBA-0002). The computational work was carried out on the (1) HPC resources at GENCI-IDRIS (Grant 2024-104137) and (2) the GRICAD infrastructure (https://gricad.univ-grenoble-alpes.fr) which is supported by the Grenoble research communities. 
\end{acknowledgement}

\begin{suppinfo}
Radial distribution functions for the  Li-F correlations, comparison of cluster distribution for short and long chain polymers, mean squared displacements, ionic current as a function of the electric field, Arrhenius plots for the self diffusion coefficients and ionic conductivity. 
\end{suppinfo}
\bibliography{bibliography}
\end{document}